\documentclass[aps,twocolumn,showpacs,amsmath,amssymb]{revtex4} 
 
\usepackage{graphicx} 
\usepackage{stmaryrd} 
\usepackage{rotating} 
\usepackage{float} 
\usepackage[scanall]{psfrag} 
\usepackage[dvips]{epsfig}
\usepackage[latin1]{inputenc}

\newcommand{\be}{\begin{eqnarray}}
\newcommand{\ee}{\end{eqnarray}}

\newcommand{\arccot}{\text{arccot}}

\newcommand{\brrm}{\protect{({\bf r},{\bf r'},\omega)}}

\newcommand{\Hb}{\protect{{\bf H}}}
\newcommand{\Eb}{\protect{{\bf E}}}

\newcommand{\Gb}{\protect{{\bf G}}}
\newcommand{\Pb}{\protect{{\bf P}}}
\newcommand{\Mb}{\protect{{\bf M}}}
\newcommand{\Sb}{\protect{{\bf S}}}
\newcommand{\Nb}{\protect{{\bf N}}}
\newcommand{\Bb}{\protect{{\bf B}}}
\newcommand{\rb}{\protect{{\bf r}}}

\newcommand{\kb}{\protect{{\bf k}}}
\newcommand{\Kb}{\protect{{\bf K}}}
\newcommand{\db}{\protect{{\bf d}}}

\renewcommand{\Im}{{\rm Im}}
\renewcommand{\Re}{{\rm Re}}
 
\begin{document} 

\title{Quantum Electrodynamics in Media with Negative Refraction
\footnote{This paper is dedicated to the 70th birthday of Herbert Walther 
whose pioneering experimental work on cavity quantum electrodynamics with 
Rydberg atoms has sharpened our understanding of the fundamentals of
light-matter interaction.}
 }
\author{J\"urgen K\"astel and Michael Fleischhauer}
\affiliation{Fachbereich Physik, Technische Universit\"{a}t 
Kaiserslautern, D-67663 Kaiserslautern, 
Germany} 
\date{\today} 

\begin{abstract} 
We consider the interaction of atoms with the quantized electromagnetic 
field in the presence of materials with negative index of refraction. 
Spontaneous emission of an atom embedded in a negative index material
is discussed. It is shown furthermore 
that the possibility of a vanishing optical path 
length between two spatially separated points provided by these materials
can lead to complete suppression of spontaneous emission of an atom
in front of a perfect mirror even if the distance between atom and mirror
is large compared to the transition wavelength. 
Two atoms put in the focal points of a lens formed by a parallel
slab of ideal negative index material are shown to exhibit perfect sub- 
and superradiance. The maximum length scale in both cases is limited only by
the propagation distance within the free-space radiative decay time.
Limitations of the predicted effects arising from absorption, finite 
transversal extension and dispersion of the material are
analyzed.
\end{abstract} 
 

 
\maketitle 
 


\section{Introduction}


Since the early days of quantum electrodynamics it is well-known and 
appreciated that
the radiative decay of an isolated atom as well as the radiative 
interaction between different atoms can be strongly affected by the
environment. As first noted by E.M. Purcell 
\cite{Purcell-PR-1946,Kleppner-1971}
the presence of conducting walls can strongly accelerate or suppress 
spontaneous emission. Inhibited emission \cite{Huket-PRL-1985},
enhanced decay \cite{Goy-PRL-1983}, and the suppression of blackbody
absorption \cite{Kleppner-PRL-1981,Walther-AnnPhys-1985} have been
observed with Rydberg atoms in cavity systems.  Alterations of the
spontaneous emission rate have also been observed near dielectric
interfaces \cite{Drexhage-Prog-Opt-1974} and in quantum-well structures
\cite{Yamamoto-OptComm-1991}. Furthermore photonic band-gap materials
with an engineered density of states of the radiation field can lead to
suppression or acceleration of spontaneous decay 
\cite{Yablonovich-PRL-1987,John-PRA-1994}. 

In this paper we discuss QED effects of single atoms and pairs of atoms
in the presence of artificial materials showing negative refraction.
Negative index materials were first predicted by V. Veselago 
\cite{Veselago-1968}, who showed that simultaneous negative values of the
dielectric permittivity $\varepsilon$ and the magnetic permeability $\mu$
imply a negative index of refraction. These so-called left-handed materials
have attracted a lot of attention, when
J. Pendry noticed that the possibility of a vanishing optical path length
between two separated points using media with a negative index of refraction
allows for a perfect lens with a resolution not limited by diffraction
\cite{Pendry-PRL-2000}. Such a lens formed by an infinite parallel slab
of lossless left-handed material of thickness $d$ collects all plane waves
from a point source on one side of the slab in a focal point on the other side.
If the refractive index of the material is $n=-1$, the distance between
the two focal points is $2d$, while the optical path between them vanishes.
We will show here that the same effect can lead to
a drastic modification of the radiative decay of a two-level atom placed 
in front of
a conducting surface (Purcell effect) and the radiative interaction 
between two atoms even if
the involved distances are large compared to the resonance wavelength.
We show that spontaneous emission from an atom with distance $2d$
from the surface of a perfect mirror can be completely suppressed for
dipole orientations in the plane of the mirror, if the space between atom
and mirror contains a slab of $n=-1$ material with thickness $d$. With this
an effect otherwise occurring only within a distance small compared to the 
transition wavelength would be observable for macroscopic distances.
We will show furthermore that two atoms put into the focal points of an ideal
Veselago-Pendry lens behave as if both would be in the same
position i.e. they show perfect Dicke-sub and superradiance \cite{Dicke54}.

After summarizing the basic properties of left-handed materials and analyzing
the conditions for their existence for the case of realistic i.e.
causal, and in general lossy magneto-dielectrics in Sec.II, we will discuss
the alteration of the spontaneous emission rate of an atom embedded in a
left-handed material in Sec.III. It will be shown that the modification
of the spontaneous emission rate due to the changed density of states is not
anymore given by the index of refraction $n$ as in dielectric materials
\cite{Nienhuis-1976},
but by the product $\mu n$, which remains positive also for lossless
negative-index materials \cite{Dung-PRA-2003}. We will then analyze the
radiative decay of a single two-level atom in front of a perfect mirror
with a layer of negative index material in Sec.IV and the radiative coupling
of two atoms in the focal points of a Veselago-Pendry lens in Sec.V. Finally
in Sec.VI we will discuss limitations due to finite absorption, a finite
transversal extension of the lens as well as due to dispersion, which
necessarily accompanies negative refraction.


\section{Electrodynamics of media with negative refractive index}


Macroscopic electrodynamics in linear, isotropic media is completely 
characterized
by the two material functions dielectric permittivity $\varepsilon$ and 
magnetic permeability $\mu$. $\varepsilon$ and $\mu$
 relate the vector of the polarization 
$\Pb$ to that of the electric field $\Eb$ and, correspondingly, the
vector of magnetization $\Mb$ to that of the magnetic field $\Bb$.
The most general expressions for $\Pb$ and $\Mb$ in linear isotropic 
magneto-dielectrics read:
\begin{equation}
\Pb(\rb,t)=\varepsilon_0\int_{-\infty}^\infty d\tau \chi_D(\rb,\tau)\Eb(\rb,t-\tau)
\label{Pkausal}
\end{equation}
and
\begin{equation}
\Mb(\rb,t)=\kappa_0\int_{-\infty}^\infty d\tau\chi_M(\rb,\tau)\Bb(\rb,t-\tau)
\label{Mkausal}
\end{equation}
where $\kappa_0=1/\mu_0$.
$\chi_D$ and $\chi_M$ are the electric and magnetic susceptibilities 
respectively.
For causality
the susceptibilities have to be zero for $\tau <0$.
The dielectric permittivity $\epsilon(\omega)$ then reads
\be
\varepsilon(\rb,\omega)=1+\int_0^\infty d\tau\chi_D(\rb,\tau)e^{i\omega\tau}.
\ee
Correspondingly the magnetic permeability $\mu=1/\kappa$ is given by
\be
\kappa(\rb,\omega)=1-\int_0^\infty d\tau\chi_M(\rb,\tau)e^{i\omega\tau}.
\ee
Causality requires that the poles of $\varepsilon(\omega)$ and
$\mu(\omega)$ are in the lower half of the complex plane. 
 $\varepsilon(\omega)$ and
$\mu(\omega)$ usually have a resonance structure in $\omega$-space, e.g.
\begin{equation}
\label{epsresonanz}
\varepsilon(\omega)=1+\frac{\omega_{Pe}^2}{\omega_{Te}^2-\omega^2-i\omega\gamma_e}
\end{equation}
and
\begin{equation}
\label{muresonanz}
\mu(\omega)=1+\frac{\omega_{Pm}^2}{\omega_{Tm}^2-\omega^2-i\omega\gamma_m}.
\end{equation}
For sufficient strength of the resonance, i.e. for $\omega_{Pe},\omega_{Pm}$ large, both $\Re[\varepsilon]$ and $\Re[\mu]$ may 
become negative for certain frequencies.

\begin{figure}[htb]
\centering
\epsfig{file=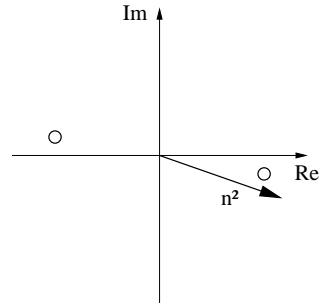,height=4cm}
\caption[Wurzelproblem]{$n^2$ for $\Re[\varepsilon],\Re[\mu]$ both being 
negative. The two possible complex square roots are indicated by circles}
\label{Wurzelproblem}
\end{figure}

\begin{figure}[htb]
\begin{center}
\epsfig{file=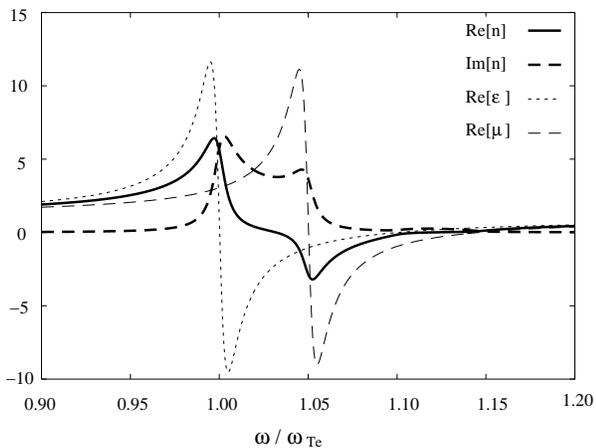,width=8cm}
\end{center}
\caption[BildMaterialneg]{$\Re[n(\omega)]$,$\Im[n(\omega)]$,
$\Re[\varepsilon(\omega)]$ and $\Re[\mu(\omega)]$ using 
eqs.(\ref{epsresonanz}) and (\ref{muresonanz}). Parameters: 
$\omega_{Pe}=\omega_{Pm}=0,46\omega_{Te}$; $\omega_{Tm}=1,05\omega_{Te}$; 
$\gamma_e=\gamma_m=0,01\omega_{Te}$.}
\label{BildMaterialneg}
\end{figure}

Suppose that at a particular frequency $\varepsilon=\mu=-1$ holds.
Then the question arises
what are the implications on the refractive index $n(\rb,\omega)$?
From the definition of $n(\rb,\omega)$ 
\begin{equation}
n(\rb,\omega)^2=\varepsilon(\rb,\omega)\cdot\mu(\rb,\omega)
\end{equation}
one might conclude
\begin{equation}
\label{obigerFall}
n=\sqrt{\varepsilon \cdot \mu} =\sqrt{(-1)\cdot(-1)} = \sqrt{1}=1.
\end{equation}
However, as pointed out by
Veselago \cite{Veselago-1968}, since $\varepsilon$ and
$\mu$ are complex functions, one has to decide which complex
root to take. Noting, that the imaginary part of
$n(\rb,\omega)$ characterizes the absorption of the medium, for a 
passive medium $\Im[n(\rb,\omega)]\ge 0$ should hold, which fixes the
root. 
As can be seen from figure \ref{Wurzelproblem} the correct choice is
\begin{equation}
\label{nkomplex}
\begin{split}
n(\rb,\omega) = & \sqrt{\left| \varepsilon(\rb,\omega)\right|\cdot\left|
\mu(\rb,\omega)\right|} \\ 
& \cdot\exp{\left[+\frac{i}{2}\left(\arccot\frac{\varepsilon_R(\rb,\omega)}
{\varepsilon_I(\rb,\omega)}+\arccot\frac{\mu_R(\rb,\omega)}
{\mu_I(\rb,\omega)}\right)\right]}. \\
\end{split}
\end{equation}
Here $\varepsilon_R,\mu_R$ and $\varepsilon_I,\mu_I$ denote real 
and imaginary parts of $\varepsilon$ and $\mu$ respectively.
%
%
%
%
%
%
With this one finds for 
the case of $\varepsilon=\mu=-1$:
\begin{eqnarray}
n & = & \lim_{\varepsilon_I,\mu \searrow 0}
\exp{\left[\frac{i}{2}\left(\arccot\frac{-1}{\varepsilon_I}
+\arccot\frac{-1}{\mu_I}\right)\right]} \nonumber\\
& = & -1\, .
\end{eqnarray}
It is easy to see from eq.(\ref{nkomplex}) that a negative real part
of the refractive index occurs if and only if \cite{Kaestel-Diplom}
\begin{eqnarray}
\pi \, \geq\, 
\arccot(\frac{\varepsilon_R}{\varepsilon_I})
+\arccot(\frac{\mu_R}{\mu_I})\,  > \, \pi/2.
\end{eqnarray}
In fig.\ref{BildMaterialneg} we have illustrated the 
frequency dependence of the index of refraction for the
single-resonance model given in (\ref{epsresonanz},\ref{muresonanz}). 
For frequencies around $\omega=1.05\omega_{Te}$
the negativity of $\Re[n]$ is clearly recognizable.
%
%
%
%
%
%
 
The example of figure \ref{BildMaterialneg} shows a strong dispersion
of the material functions $\varepsilon(\omega)$ and $\mu(\omega)$.
In fact, as pointed out already by Veselago, this is a general property of
negative-index materials. 
Considering the energy of the electromagnetic field in a non-dispersive medium
\begin{equation}
\label{Energiealt}
w=\varepsilon\Eb^2+\mu\Hb^2
\end{equation}
one recognizes that negative values of $\varepsilon$ and $\mu$ would lead to
a negative energy.
Therefore a negative index of refraction is necessarily associated 
with dispersion, in which case 
the energy of the electromagnetic field reads
\begin{equation}
w=\Re\left[\frac{d(\omega\varepsilon)}{d\omega}\right]\Eb^2
+\Re\left[\frac{d(\omega\mu)}{d\omega}\right]\Hb^2
\end{equation}
wich is positive even for negative $\varepsilon$ and $\mu$ if 
the dispersion is normal and sufficiently large, such that
\begin{equation}
\Re\left[\frac{d(\omega\varepsilon)}{d\omega}\right] \geq 0,\quad\textrm{and}
\quad
\Re\left[\frac{d(\omega\mu)}{d\omega}\right] \geq 0.
\end{equation}

\begin{figure}[H]
\begin{center}
\epsfig{file=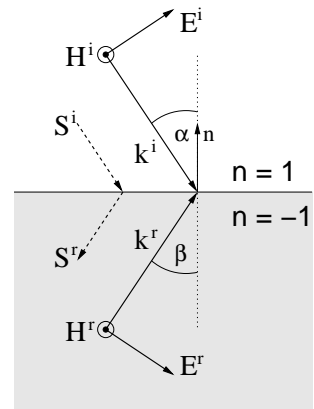,width=4cm}
\end{center}
\caption{Boundary between positive and negative refraction. 
In a material with a 
negative refractive index, the refracted wave goes to negative angles.}
\label{Randbedingungen}
\end{figure}

In the following we want to discuss some of the peculiar aspects of
light propagation in negative-index materials.
Making use of the boundary conditions between media with positive
and negative refractive indexes, namely ${\bf E^\perp,H^\perp}$ being 
continuous as well as ${\bf D^\parallel,B^\parallel}$, one finds that
an incident plane wave is refracted to the same side of the normal
as shown in figure \ref{Randbedingungen}. This behavior is fully
consistent with Snells law
\begin{equation}
\frac{n_2}{n_1} = \frac{\sin(\alpha)}{\sin(\beta)}.
\end{equation}
One striking feature of negative refraction is, that the wave vector of
the refracted wave $\kb^r$ points backward which is due to 
conservation of momentum parallel to the surface. 
As a result the vectors ${\bf k,E,H}$ form a left-handed tripod 
instead of the ususal right-handed one.
Materials with
a negative index of refraction are therefore also 
called left-handed media (LHM).

On the other hand the Poynting vector $\Sb=\Eb\times\Hb$ clearly forms a 
right-handed tripod with $\Eb$ and $\Hb$ and therefore points in the 
correct direction, namely away from the surface, as it should be due to 
conservation of energy (dashed arrows in fig. \ref{Randbedingungen}).

%
%
%
%
%
%

\begin{figure}[ht]
\begin{center}
\epsfig{file=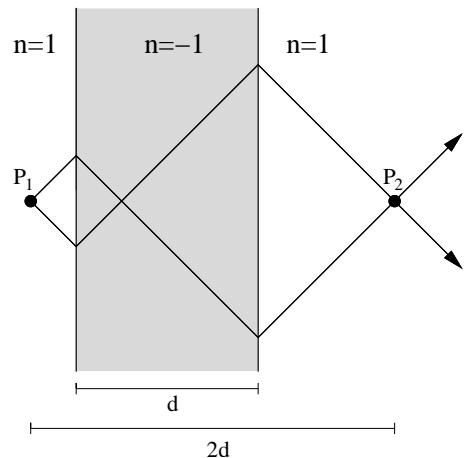,height=6.0cm}
\end{center}
\caption[Pendrylinse]{Perfect lens formed by a infinitely extended slab of a 
LHM. The optical length between the foci
$P_1$ and $P_2$ is zero.}
\label{Abb1}
\end{figure}

Besides strong influences on Doppler and Cherenkov effects \cite{Veselago-1968}, 
the most prominent effect of the negative refraction is probably 
the so-called perfect lens formed by a slab of a LHM. 
It was Veselago \cite{Veselago-1968} who, when first 
studying the properties of lenses formed by a left-handed material, 
found that an infinitely extended slab
of a LHM  collects all plane waves coming from a point
source not too far away from the surface in a focal point on the 
other side of the slab (fig. \ref{Abb1}).
%
%
%
%
%
%
For a slab of thickness $d$ the distance between the two focal 
points is $d(1-n)$, where $n$ is the 
refractive index of the LHM. Noting that the optical path-length between 
points $P_1$ and $P_2$
\begin{equation}
l^{\text{opt.}}=(d(1-n)-d)+nd = 0 ,
\end{equation}
the perfection of the lens becomes clear. The lens simulates point $P_2$ 
to be at the same spatial position
as point $P_1$. Traveling a zero path no information about the source
can get lost, including that contained in the evanescent waves. 
Therefore the perfect lens 
allows an unlimited resolution of the source \cite{Pendry-PRL-2000}.


\section{Spontaneous emission of an atom embedded in a
medium with negative refraction}

\begin{figure}[H]
\begin{center}
\epsfig{file=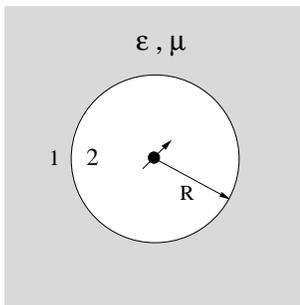,height=4cm}
\end{center}
\caption[real-cavity]{Real-cavity model: the atom is located at the center 
of an empty cavity surrounded by the dielectric body. }
\label{realcavfig}
\end{figure}

It is a well known fact, that the natural linewidth $\Gamma$ of a 
dipole allowed transition is not an 
intrinsic feature of an atom, but depends on the local environment.
Based on an analysis of the density of states of the radiation field
Nienhuis and Alkemade predicted for an atom embedded in a homogeneous
transparent dielectric with refractive index $n$ \cite{Nienhuis-1976}
\begin{equation}
\Gamma = \Gamma_0\, n, \label{Nienhuis}
\end{equation}
where $\Gamma_0$ is the free-space decay rate:
\begin{equation}
\Gamma_0=\frac{d^2 \omega_A^3}{3\pi\hbar \varepsilon_0 c^3}
\end{equation}
with $d$ being the dipole moment and $\omega_A$ the atomic transition 
frequency. It was seen later on that eq.(\ref{Nienhuis}) does 
not give the correct behavior of $\Gamma$ since 
the macroscopic description of the 
surrounding medium fails in the immediate
environment of the probe atom. To correct this in leading order of the
medium density, local-field corrections need to be included.
Several models have been established for this. The one that best describes 
a substitutive probe atom in a cubic-lattice host  
is the so-called real-cavity-model \cite{Glauber} in which
\be
\Gamma=n\Gamma_0{\mathcal L}_\text{real}^2.
\label{GlauberLewenstein}
\ee
Here  
${\mathcal L}_\text{real}={3\varepsilon}/{(2\varepsilon+1)}$ 
is the Glauber-Lewenstein factor 
which accounts for near field effects. 
This model assumes the atom to be located at the center 
of a small empty cavity surrounded by the dielectric body which 
is treated macroscopically (see fig. \ref{realcavfig}).
%
%
%
%
%
%

When considering atoms embedded in negative index materials 
eqs.(\ref{Nienhuis}) and (\ref{GlauberLewenstein}) are
obviously not correct, since $\Gamma$ would become
negative in this case. This is because these expressions are derived 
only for dielectric surroundings \cite{Glauber,Knoell},
where the contribution of the magnetic dipoles of the material was neglected.

Following Fermis golden rule the rate of spontaneous emission of an electric
dipole transition is given by the imaginary part of the retarded Greensfunction
$\Gb$ of the electric field at the position $\rb_A$ of the atom and at the 
transition frequency $\omega_A$ \cite{Dung-PRA-2003}
\begin{equation}
\label{Gamma}
\Gamma = \frac{2\omega_A^2d_id_j}{\hbar\varepsilon_0c^2}
\Im \left[G_{ij} (\rb_A,\rb_A,\omega_A)\right].
\end{equation}
Here $d_i$ is the Cartesian $i$th-component of the 
dipole moment. The influence of the surrounding on the mode structure 
is contained in the Greenstensor $\Gb$.
Thus in order to include the effect of the magnetic dipoles of the
left-handed medium, the Greenstensor needs to be calculated for the case
of a magneto-dielectrics. Thus the solution of the equation
\begin{equation}
\label{Bestimmung}
\begin{split}
& \left(\nabla_\rb\times\left[ \kappa(\rb,\omega)\nabla_\rb\times\right] 
-\frac{\omega^2}{c^2}\varepsilon(\rb,\omega)\right) \Gb\brrm \\
& \qquad\qquad =\delta(\rb-\rb'){\bf 1}
\end{split}
\end{equation}
needs to be determined for given boundary conditions.
Note the term $\kappa(\rb,\omega)=1/\mu(\rb,\omega)$ which is absent
in the pure dielectric case \cite{Knoell}.

Since within the Glauber-Lewenstein model 
the atom is located in free space, the solution of (\ref{Bestimmung}) for the 
real-cavity (fig. \ref{realcavfig}) can be expressed as a 
sum of the free space Greensfunction and a scattering term accounting 
for the boundary to the magneto-dielectric:
\begin{equation}
\label{Aufteilung}
\Gb(\rb,\rb')=\Gb^{vac}(\rb,\rb')+\Gb^s(\rb,\rb')  \qquad \left|\rb\right|,\left|\rb'\right|<R
\end{equation}
with
\begin{equation}
\Im \left[\Gb^{vac}_{ij}(\rb,\rb,\omega)\right] = \frac{k}{6\pi}\delta_{ij},
\qquad k=\omega/c
\end{equation}
and
\begin{equation}
\label{Greenrealcav}
\begin{split}
\Gb^s(\rb,\rb') =
& \frac{ik}{4\pi}\sum_{n=1}^\infty \sum_{m=0}^n \sum_{l=e,o}(2-\delta_{0m})\frac{2n+1}{n(n+1)}\frac{(n-m)!}{(n+m)!} \\
& \cdot\Bigl[ {\mathcal C}_M^n(k) \Mb_{lmn}(\rb,k)\Mb_{lmn}(\rb',k) \\
& +{\mathcal C}_N^n(k) \Nb_{lmn}(\rb,k)\Nb_{lmn}(\rb',k) \Bigr].
\end{split}
\end{equation}
Due to the symmetry of the geometry the scattering term can be 
expanded in a series of vector Bessel functions:
\begin{eqnarray}
\Mb_{lmn}(\rb,k) &=& \nabla\times [\psi_{lmn}(\rb,k)\rb],\\
\Nb_{lmn}(\rb,k) &=& 
\frac{1}{k}\nabla\times\nabla\times [\psi_{lmn}(\rb,k)\rb],\\
\psi_{lmn}(\rb,k) &=& j_n(kr)P_n^m(\cos\theta) f_l(m\phi),
\end{eqnarray}
$f_l$ has the meaning of a cosine function for even $l$ and of a sine 
function for $l$ being odd. The $j_n$ are spherical Bessel functions 
of the first kind and the $P_n^m$ associated Legendre polynomials. 
The rather lengthy expansion factors ${\mathcal C}_N^n(k)$
and ${\mathcal C}_M^n(k)$ are given in \cite{Li}.

In the limit $\rb,\rb'\rightarrow 0$ all terms but 
that with ${\mathcal C}_N^1$ become zero. The rate of spontaneous 
emission of a 2-level atom embedded in a medium of arbitrary 
$\varepsilon$ and $\mu$ then reads
\begin{equation}
\Gamma = \Gamma_0 \Bigl(1+\Re \left[{\mathcal C}_N^1(\omega_A)\right]\Bigr)
\label{Gammarealcav}
\end{equation}
with
\begin{equation}
\begin{split}
& \qquad\qquad{\mathcal C}_N^1(\omega_A) = \\ 
& e^{i\varrho}\left[i+\varrho(n+1)-i\varrho^2 n(n+1)\frac{\mu-n}{\mu-n^2}-\varrho^3n^2\frac{\mu-n}{\mu-n^2}\right] \\
& \cdot\left[i\varrho^2n\left(\cos\varrho+in\frac{\mu-1}{\mu-n^2}\sin\varrho\right)-\varrho(\cos\varrho+in\sin\varrho)\right. \\
& \qquad \left. \sin\varrho + \varrho^3(\mu\cos\varrho-in\sin\varrho)\frac{n^2}{\mu-n^2}\right]^{-1}.
\end{split}
\end{equation}
Here $\varrho=\frac{R\omega_A}{c}$ is the normalized radius of the cavity.
This function can be shown to be strictly positive, as it should be 
for the rate of spontaneous emission. In the limit of vanishing
imaginary parts of $\mu$, $\varepsilon$, and $n$, eq.(\ref{Gammarealcav}) 
reduces to
\begin{equation}
\Gamma =n \mu\, \Gamma_0 \, \left(\frac{3 \varepsilon}
{2\varepsilon+1}\right)^2
\label{GlauberLewensteinneu}
\end{equation}
which is the sought generalization of the formula of Glauber and 
Lewenstein (\ref{GlauberLewenstein}) for pure dielectrics 
to the case of lossless but otherwise arbitrary magneto-dielectric media. 
For $\mu=1$ equation (\ref{GlauberLewensteinneu}) reduces to the 
dielectric case.

\begin{figure}[htb]
\begin{center}
\epsfig{file=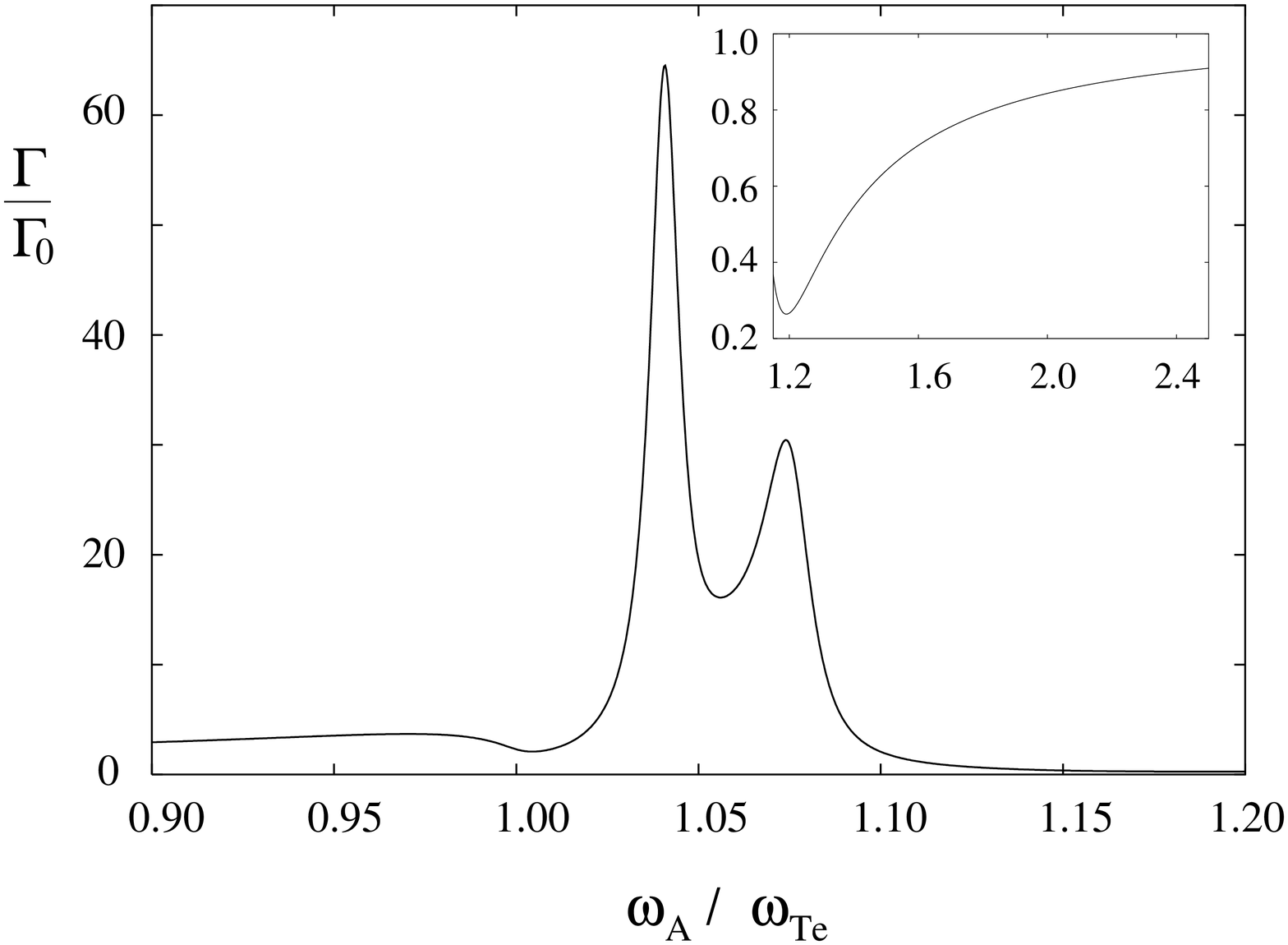,width=8cm}
\end{center}
\caption{Rate of spontaneous emission for the real-cavity 
model for the resonant functions (\ref{epsresonanz}) and 
(\ref{muresonanz}). $\omega_{Pe}=\omega_{Pm}=0,46\omega_{Te}$, 
$\omega_{Tm}=1,05\omega_{Te}$ and $\gamma_e=\gamma_m=0.01\omega_{Te}$}
\label{gammagraph}
\end{figure}

The influence of the LHM on the rate of spontaneous 
emission for a resonance model using the example 
of eqs.(\ref{epsresonanz}) and (\ref{muresonanz}) is 
shown in figure \ref{gammagraph}.
%
%
%
%
%
%
Because of the surrounding medium the natural linewidth in the vicinity of
the resonance 
can be either strongly enhanced or suppressed.


\section{Suppression of spontaneous emission of an atom in front of a mirror:
modified Purcell effect}


\label{pmirror}

\begin{figure}[H] 
  \begin{center} 
    \includegraphics[width=4cm]{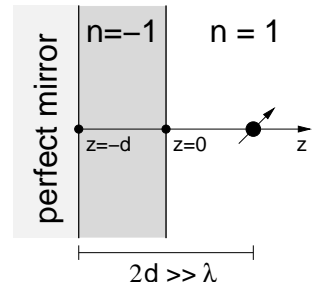} 
    \caption{Atom in front of a LHM attached to a mirror. The optical 
length between atom and mirror is zero.
The spatial regions $z>0$, $-d\le z\le 0$, and $z<-d$
are denoted by the numbers $0,1,2$ respectively.}
    \label{perfectmirror} 
  \end{center} 
\end{figure}

In this section the modification of the Purcell effect, i.e. the
suppression of the spontaneous emission of an atom in front of a mirror
by a medium with negative refraction will be discussed. 
For this we consider the setup shown in fig. \ref{perfectmirror}. 
The atom is placed at a distance $2d$ from the surface of the perfect
mirror and the space between atom and mirror is filled half by vacuum
and half by a medium with $n=-1$.
%
%
%
%
%
%
In the absence
of the medium, the emission rate of the atom is significantly affected 
only if the distance between atom and mirror is small compared to
the transition wavelength \cite{Morawitz}. In this case the radiative
decay for a transition with dipole moment parallel to the plane
of the mirror vanishes, while that for an orthogonal dipole moment is enhanced
by a factor 2. 
Since in the presence of the negative index material as in 
fig.\ref{perfectmirror},
the optical length of the path from the atom to the mirror equals zero,
the question arises whether the LHM leads to properties 
comparable
to the case of the atom sitting on the mirror surface.
To obtain an answer to this we note, that the rate of spontaneous 
emission of an atom 
in a linear, isotropic but otherwise arbitrary environment is
given by (\ref{Gamma}). Thus we only need to calculate the Greensfunction
corresponding to the specific set-up.

The retarded Greensfunction corresponding to a slab 
with a homogeneous and linear magneto-dielectric medium
can be calculated
by a plane wave decomposition. Following \cite{Tsang85} one finds for the two
positions $\rb$ and $\rb^\prime$ in vacuum on the same side of the slab
\begin{eqnarray}
&&\Gb^{00}(\rb,\rb',\omega) =  
\frac{i}{8\pi^2}\int d^2 k_\perp \frac{1}{k_z} \Bigl[
\label{Green00}\\
&& \qquad 
 \bigl( R^{\rm TE}\hat {\bf e}(k_z) e^{i\kb\cdot\rb} 
+ \hat {\bf e}(-k_z) e^{i\Kb\cdot\rb} \bigr) 
\circ \hat {\bf e}(-k_z) e^{-i\Kb\cdot\rb'} \nonumber\\
&&\quad + \bigl(R^{\rm TM} \hat {\bf h}(k_z) e^{i\kb\cdot\rb} 
+ \hat {\bf h}(-k_z) e^{i\Kb\cdot\rb} \bigr) \circ
\hat {\bf h}(-k_z) e^{-i\Kb\rb'} \Bigr],\nonumber
\end{eqnarray}
where $z\le z^\prime$ has been assumed. 
Since it is needed later, we also give the Greensfunction 
for $\rb$ and $\rb^\prime$ being in vacuum on different sides 
of the slab
\begin{eqnarray}
&&\Gb^{20}(\rb,\rb',\omega)=  
\frac{i}{8\pi^2}\int d^2 k_\perp \frac{1}{k_z} \Bigl[
\nonumber\\
&&\qquad  T^{\rm TE}\hat {\bf e}(-k_z) e^{i\Kb\cdot\rb} 
\circ \hat {\bf e}(-k_z) e^{-i\Kb\cdot\rb'} \label{Green20}\\
&&\quad + T^{\rm TM} \hat {\bf h}(-k_z) e^{i\Kb\cdot\rb} \circ
\hat {\bf h}(-k_z) e^{-i\Kb\rb'} \Bigr].\nonumber 
\end{eqnarray}
The superscripts $0,1,2$ at the Greensfunctions 
denote the zones of positions $\rb$ and $\rb'$: 
$z>0$, $-d\le z\le 0$, and $z<-d$
respectively.
For later convenience $\Gb^{20}$ is given under the 
assumption of medium 2 being vacuum.
We here have used the definitions $k^2 ={\omega^2}/{c^2}$, 
$k_z =\sqrt{k^2-k_\perp^2}$ and
$d^2k_\perp = dk_xdk_y$. Furthermore
$\Kb \equiv k_x\hat {\bf x}+k_y\hat {\bf y}-k_z\hat {\bf z}$ and we have
introduced the orthogonal unit vectors 
$\hat {\bf e} = {\kb\times\hat {\bf z}}/{\left|\kb\times\hat {\bf z}\right|}$
and $\hat {\bf h} = {p}\hat {\bf e}\times \kb/\left|k\right|$, where $p=1$ for
a normal medium and $p=-1$ for a LHM.
$R^{\rm TE}, R^{\rm TM}$ and $T^{\rm TE}, T^{\rm TM}$ are the reflection and
transmission functions of the 3-layer medium for transverse 
electric and transverse
magnetic modes. They read
\begin{eqnarray}
\label{RTE}
R^{\rm TE} &=& \frac{R_{01}+R_{12}e^{i2k_{1z}d}}{1+R_{01}R_{12}e^{i2k_{1z}d}}, \\
\label{RTM}
R^{\rm TM} &=& \frac{S_{01}+S_{12}e^{i2k_{1z}d}}{1+S_{01}S_{12}e^{i2k_{1z}d}},
\end{eqnarray}
and correspondingly
\begin{equation}
\label{TTE}
T^{\rm TE} = \frac{2\mu k_z}{\mu k_z+k_{1z}}\frac{1+R_{12}}
{1+R_{01}R_{12}e^{i2k_{1z}d}}e^{i(k_{1z}-k_z)d},
\end{equation}
\begin{equation}
\label{TTM}
T^{\rm TM} = \frac{2\varepsilon k_z}{\varepsilon k_z+k_{1z}}
\frac{1+S_{12}}{1+S_{01}S_{12}e^{i2k_{1z}d}}e^{i(k_{1z}-k_z)d}.
\end{equation}
Here $k_{1z}=\sqrt{k_1^2-k_\perp^2}$ and 
$k_1^2 = \varepsilon(\omega)\mu(\omega)  \omega^2/c^2$.
$R_{ij}$ and $S_{ij}$ are the reflection coefficients at the boundaries bet\-ween
media $i$ and $j$ for TE and TM modes respectively.
\begin{equation}
\label{RS}
R_{ij} = \frac{\mu_jk_{iz}-\mu_ik_{jz}}{\mu_jk_{iz}+\mu_ik_{jz}},
\qquad S_{ij} = \frac{\varepsilon_jk_{iz}-\varepsilon_ik_{jz}}
{\varepsilon_jk_{iz}+\varepsilon_ik_{jz}}.
\end{equation}

\begin{figure}[htb]
  \begin{center}
  \epsfig{file=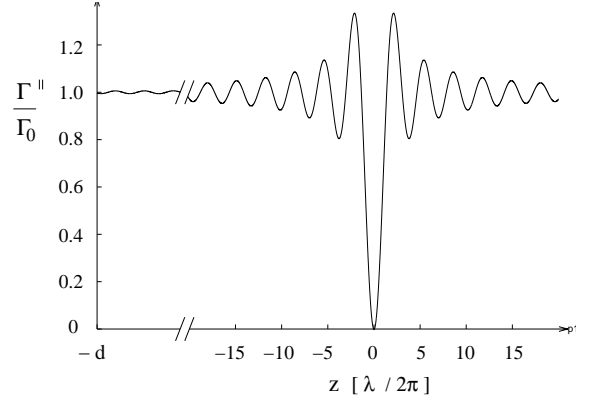,width=7.5cm}
     \caption{Spatial dependence of the normalized rate of 
spontaneous emission $\Gamma^\parallel(z)/\Gamma_0$ for dipole parallel
to mirror surface. 
$z$ is the spatial shift in z-direction of the atom out of its focus; 
$d$ is the distance from the focus to the surface of the LHM.}
     \label{atomvorspiegel}
  \end{center}
\end{figure}

Setting $R_{12}=-1$ and $S_{12}=1$ to account for the perfect mirror, we get 
the natural linewidth by substituting the corresponding result for
$\Gb^{00}$ into
\begin{equation}
\label{Gammamirror}
\Gamma = \frac{2\omega_A^2d_id_j}{\hbar\varepsilon_0c^2}\Im \left[\Gb_{ij}^{00} (\rb_A,\rb_A,\omega_A)\right].
\end{equation}
The result is shown in figure \ref{atomvorspiegel} for the case of the 
atomic dipole moment being parallel to the surface of the mirror.
The thickness of the LHM is set to $d=100\frac{\lambda}{2\pi}$. When the atom 
is put at a distance $d$ before the LHM, which we want to denote as focal
point, the rate of spontaneous
emission is completely suppressed:
\begin{equation}
\Gamma^\parallel_\textrm{focus}=0.
\end{equation}
The spatial dependence 
of the linewidth shown in figure \ref{atomvorspiegel} is the same as for an 
atom in front of a mirror located at $z=0$ without LHM \cite{Morawitz}.
%
%
%
%
%
%
For atomic dipoles with orthogonal orientation to the mirror, 
the spatial dependence is also the same as in the case of no LHM but the 
mirror being at $z=0$ position (see \cite{Morawitz}). In the focus this leads
to an enhancement of the decay rate by a factor of 2:
\be
\Gamma^\perp_\text{focus}=2\Gamma_0
\ee
The behavior shown in fig.\ref{atomvorspiegel} can be understood in the
following way: The combination of layers of vacuum ($n=+1$) and of
negative index material ($n=-1$) with equal thickness $d$ makes the
space between atom and mirror to appear of zero (optical) length.
Thus the atom in the focal point is equivalent to the atom being on the
mirror surface.
This result suggests the possibility to
experimentally study spontaneous emission suppression
of atoms near a mirror without the necessity of actually putting the atoms
on the surface.


\section{Sub- and Superradiance over macroscopic distances}


\label{plens}

\begin{figure}[ht] 
  \begin{center} 
    \includegraphics[width=5cm]{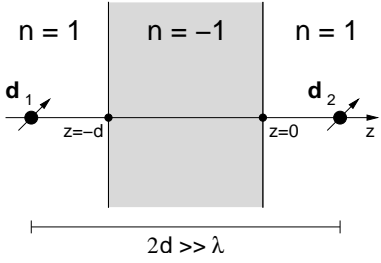} 
    \caption{two atoms put into the focal points of a
Veselago-Pendry lens with $n=-1$. Focal points are all pairs of positions
at the two sides of the slab with distance $2d$.
The spatial regions $z>0$ (vacuum), $-d\le z\le 0$ (LHM), and $z<-d$ (vacuum)
are denoted by the numbers $0,1,2$ respectively.}
    \label{perfectlens} 
  \end{center} 
\end{figure}

The observation of the last section, that a combination of a layer of
positive and negative refraction can make a spatial volume to
appear to have vanishing optical thickness, suggests a different interesting
application. If two atoms are put in the focal points of a Veselago-Pendry 
lens, as indicated in fig.\ref{perfectlens}, they should show a 
radiative coupling with a strength as if they would be
at the same position in space.
%
%
%
%
%
%

We now want to analyze this situation in detail. For this we start with
the interaction Hamiltonian of two atoms at positions $\rb_1$ and $\rb_2$
with the quantized electric field $\hat \Eb$ in dipole and rotating-wave
approximation:
\begin{equation}
H_{\text{WW}} = -\hat\Eb(\rb_1)\hat\db_1 -\hat\Eb(\rb_2)\hat\db_2 .
\end{equation}
Eliminating the electromagnetic field using the usual Born-Markov
approximations leads to a two-atom Liouville equation of the form
\begin{equation}
\label{Liouville}
\begin{split}
\dot\varrho = & -\frac{i}{\hbar}\left[\hat H_0,\varrho\right] \\
& -\sum_{k,l=1}^2\frac{\Gamma(\rb_k,\rb_l)}{2}(\hat\sigma^{\dagger}_l
\hat\sigma_k\varrho+\varrho\hat\sigma^{\dagger}_l\hat\sigma_k-2\hat
\sigma_k\varrho\hat\sigma^{\dagger}_l) \\
& +\frac{i}{\hbar}\sum_{k,l=1}^2\delta\omega(\rb_k,\rb_l)\left[\hat
\sigma^{\dagger}_l\hat\sigma_k,\varrho\right],
\end{split}
\end{equation}
where $\hat\sigma_l=|1\rangle_{ll}\langle 2|$ is the atomic flip operator
of the $l$th atom from the lower state $|1\rangle$ to the upper 
state $|2\rangle$. 
The second and third term in (\ref{Liouville}) 
describe the spontaneous emission and Lamb-shift
of the two individual atoms
with decay rates $\Gamma(\rb_i,\rb_i)$ and respective level shifts 
$\delta\omega(\rb_i,\rb_i)$, $(i=1,2)$. 
However they also contain terms describing the radiative interaction between
the atoms containing a dissipative cross coupling proportional to 
$\Gamma(\rb_1,\rb_2)$ and a conditional level shift proportional to 
$\delta\omega(\rb_1,\rb_2)$.
The single-atom and cross-coupling rates are given by an expression
similar to (\ref{Gamma}):
\begin{equation}
\label{Gamma2}
\Gamma(\rb_k,\rb_l) = \frac{2\omega_A^2d_id_j}{\hbar\varepsilon_0c^2}
\Im \left[G_{ij} (\rb_k,\rb_l,\omega_A)\right].
\end{equation}
The level shifts read
\begin{equation}
\label{delta-omega}
\delta\omega(\rb_k,\rb_l) = \frac{d_id_j}{\hbar\pi \varepsilon_0}
{\cal P} \int_0^\infty\!\!{\rm d}\omega\, \frac{\omega^2}{c^2} 
\frac{\Im \left[G_{ij} (\rb_k,\rb_l,\omega)\right]}{\omega-\omega_A}.
\end{equation}
The single-atom Lamb shift $\delta\omega(\rb_i,\rb_i)$ is not accurately 
described within the present theory and will be ignored in the following.
One recognizes from (\ref{Gamma2}) and (\ref{delta-omega}) 
that the radiative interaction between the two atoms is
determined by the imaginary part of the retarded Greensfunction between the
positions $\rb_1$ and $\rb_2$ of the two atoms. In free space the value of 
$\Gb$ rapidly decreases if the relative distance $|\rb_1-\rb_2|$ becomes
larger than the transition wavelength $\lambda$. Consequently the 
radiative interaction is negligible except for very small distances. 
As will be shown now this situation changes if the atoms
are put into the focal points of a Veselago-Pendry lens. 

In order to see the effect of the radiative coupling it is convenient to
use as a basis for the two-atom system besides the total ground state
$|11\rangle$ and the doubly-excited state $|22\rangle$ the symmetric and
antisymmetric combinations of one atom being excited ($|2\rangle$) and
one atom being in its ground state ($|1\rangle$):  
\begin{eqnarray}
|s\rangle &\equiv & \frac{1}{\sqrt{2}}\Bigl(|12\rangle + |21\rangle\Bigr),\\
|a\rangle &\equiv & \frac{1}{\sqrt{2}}\Bigl(|12\rangle - |21\rangle\Bigr).
\end{eqnarray}
In terms of these basis states we arrive at the following density-matrix 
equation
\begin{eqnarray}
\dot\rho_{22} &=& -2 \Gamma_{11} \rho_{22},\\
\dot\rho_{\rm ss} &=& -(\Gamma_{11}+\Gamma_{12}) \rho_{\rm ss}
+(\Gamma_{11}+\Gamma_{12}) \rho_{22}
,\label{symm}\\
\dot\rho_{\rm {\rm aa}} &=& -(\Gamma_{11}-\Gamma_{12}) \rho_{{\rm aa}}
+(\Gamma_{11}-\Gamma_{12}) \rho_{22},
\label{asymm}\\
\dot\rho_{11} &=& + (\Gamma_{11}+\Gamma_{12})\rho_{\rm ss}
+(\Gamma_{11}-\Gamma_{12})\rho_{{\rm aa}},
\end{eqnarray}
where we have disregarded the level shifts and $\Gamma_{ij}$ is a short
notation for $\Gamma(\rb_i,\rb_j)$. One recognizes from the above equations
that the decay channels through the symmetric and antisymmetric superpositions
differ by the cross-coupling contribution $\pm\Gamma_{12}$.  
If the two atoms are in free space at the same point 
$\Gamma_{12}=\Gamma_{11}=\Gamma_{22}$. In this case the antisymmetric state 
does not decay at all, while the symmetric one decays with twice the
single-atom decay rate. This is the situation of Dicke sub- and superradiance
\cite{Dicke54}.

\begin{figure}[H]
\begin{center}
\epsfig{file=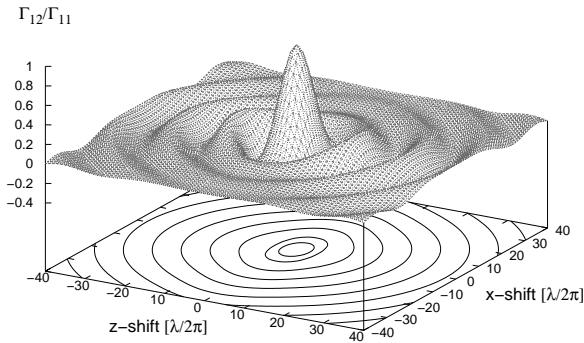,width=8cm}
\end{center}
\caption[Ortsabhängigkeit parallel]{$\Gamma_{12}/\Gamma_{11}$ as 
function of the spatial shift, 
parallel '$x$' and orthogonal '$z$' to the surface of the LHM, of one atom
out of the focus of the other one. In the focal point $(0,0)$ perfect sub- and
superradiance is obtained. Here the dipoles of both atoms are oriented 
in $x$-direction}
\label{Ortparallel}
\end{figure}

Let us now calculate the rates $\Gamma_{11}, \Gamma_{22}$ and $\Gamma_{12}$
i.e. the imaginary part of the Greensfunction
for the situation of fig.\ref{perfectlens}.
Since at the boundary between vacuum ($n=+1$) and
LHM ($n=-1$) there is no reflection, i.e. $R^{TE}=R^{TM}=0$ one finds
for the case of both positions 
being on the left side of the lens (region ``0'')
\begin{equation}
\begin{split}
\Im \left[G^{00}_{\mu\mu}(\rb,\rb)\right] & =\Im\frac{i}{8\pi^2}\int_0^\infty dk_\perp \frac{k_\perp}{\sqrt{k^2-k_\perp^2}}\cdot \\
& \left((1+R^{TE}e^{i\sqrt{k^2-k_\perp^2}d})\pi\right. \\
& \left.+R^{TM}e^{i\sqrt{k^2-k_\perp^2}d}\pi(\frac{k_\perp^2}{k^2}-1)+\pi(1-\frac{k_\perp^2}{k^2})\right) \\
= & \frac{k}{6\pi}.
\end{split}
\end{equation}
The same result holds of course for both positions being on the 
right side of the lens (region ``2''). A corresponding calculation
for $\rb$ being on the left side (region ``0'') and $\rb^\prime$
being the other focal point of the lens $\rb+2 d{\bf e}_z$ 
yields 
\begin{equation}
\begin{split}
\Im \left[\Gb^{20}_{11}(\rb\right. &,\left.\rb+2 d{\bf e}_z)\right] =\Im\frac{i}{8\pi^2}\int_0^\infty dk_\perp \frac{k_\perp}{\sqrt{k^2-k_\perp^2}} \\
& \cdot e^{i2\sqrt{k^2-k_\perp^2}d}\cdot\left(T^{TE}\pi+T^{TM}\pi(1-\frac{k_\perp^2}{k^2})\right) \\
= & \frac{k}{6\pi},
\end{split}
\end{equation}
where $T^{TE}=T^{TM}=e^{i(k_{1z}-k_z)d}$ and $k_{1z}=-k_z$ 
(\ref{TTE},\ref{TTM}) have been used. Thus we recognize that the imaginary
part of the Greensfunction between the two focal points is identical to
that at the same position. As a consequence $\Gamma_{12}=\Gamma_{11}$ 
and there is perfect sub- and superradiance, despite the fact that
the distance between the focal points can be much larger than the
resonance wavelength. Fig.\ref{Ortparallel} illustrates the dependence
of the ratio $\Gamma_{12}/\Gamma_{11}$ on the spatial
displacement of the second atom from the focal point of the first.
%
%
%
%
%
%

One recognizes from fig. \ref{Ortparallel} that for the imaginary part of
the Greensfunction of the ideal Veselago-Pendry lens holds
\begin{equation}
\Im[\Gb(\rb-2d{\bf e}_z,\rb,\omega)]=\Im[\Gb(\rb,\rb,\omega)].\label{imag-G}
\end{equation}
It should be noted here that a relation similar to (\ref{imag-G}) does not
exist for the real part of the Greensfunction. If this would be true
the electric field pattern at the two focal points would be identical in
violation of Maxwells equations. It should also be noted that relation
(\ref{imag-G}) holds only within a certain range of frequencies $\omega$ due to
the necessarily dispersive nature of the LHM. The limitations arising from 
this will be discussed in the following section.


\section{Limitations}


In sections \ref{pmirror} and \ref{plens} two systems involving 
perfect left-handed materials were discussed.
We here turn to the question what are the limitations of the
observed effects under more realistic conditions, i.e. when taking into
account absorption losses, a finite transversal extension of the LHM slab, and
dispersion of the medium.


\subsection{Absorbing LHMs}


First of all to describe a more realistic LHM one has to take into account
absorption. This can easily be done by substituting the refractive 
index $n=-1$
of the perfect LHM by $n=-1+in_I$, i.e. by adding an imaginary part. 

\begin{figure}[htb] 
  \begin{center} 
    \epsfig{file=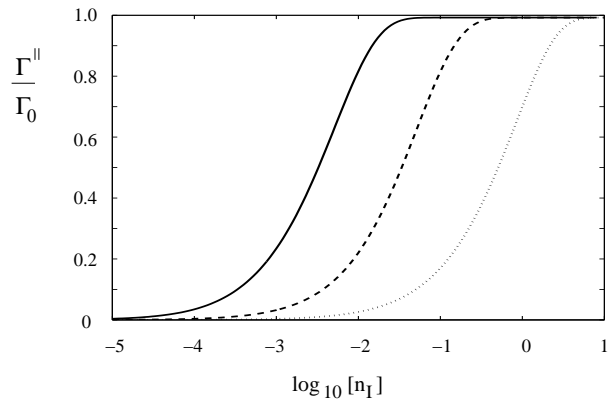,width=8cm}
    \caption{$\Gamma^\parallel/\Gamma_{0}$ as function of imaginary
part of refractive index $n_I$ for Re$[n]=-1$ and different thickness $d$
of the lens, $d = 100 \lambda/2\pi$ (solid line), 
$d= 10 \lambda/2\pi$ (dashed), and
$d=1 \lambda/2\pi$ (dotted).}
    \label{absabhmirror} 
  \end{center} 
\end{figure}

For the mirror-system of sect.IV, figure \ref{absabhmirror} shows the 
dependence of $\Gamma^\parallel/\Gamma_{0}$ 
on the absorption coefficient $n_I$ for different thicknesses of the LHM.
%
%
%
%
%
%
As can be seen, the suppression of the spontaneous emission, for atomic dipoles
parallel to the mirror, decreases with increasing absorption as expected. 
The sensitivity to absorption is approximately exponential 
in $n_I d$.

\begin{figure}[htb] 
  \begin{center} 
    \epsfig{file=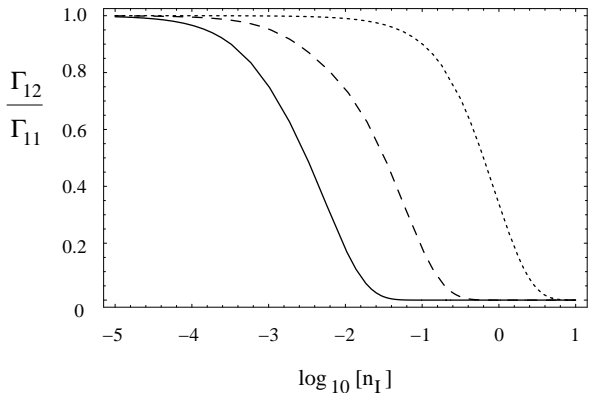,width=8cm}
    \caption{$\Gamma_{12}/\Gamma_{11}$ as function of imaginary
part of refractive index $n_I$ for Re$[n]=-1$ for different thicknesses $d$
of the lens, $d = 100 \lambda/2\pi$ (solid line), 
$d= 10 \lambda/2\pi$ (dashed), and
$d=1 \lambda/2\pi$ (dotted).}
    \label{absabh} 
  \end{center} 
\end{figure}

In figure \ref{absabh} the same is shown for the system with the lens. 
%
%
%
As expected, $\Gamma_{12}/\Gamma_{11}$, and therefore the effect 
of the sub/superradiance reduces with increasing absorption
coefficients $n_I$. The dependence on the thickness of the lens is again 
exponential.


\subsection{Finite transverse extension of the LHM}

\begin{figure}[H]
\begin{center}
\epsfig{file=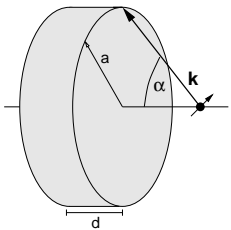,width=3cm}
\end{center}
\caption{Finite extension of the LHM in transverse direction.}
\label{Pendry3D}
\end{figure}

In experimental implementations the slab of LHM will always have a finite
transversal extension. We therefore analyze here
the dependence of the LHM-induced effects on the transversal
radius of the medium.
The thickness of the LHM is denoted by $d$, the transverse
extension by $a$ (fig. \ref{Pendry3D}).
%
%
%
%
%
%
Because of the finite radius the 1-D character of the geometry is no longer 
given and
the Greensfunction cannot be calculated analytically. A numerical
solution for the Greensfunction is also very difficult. 
Noting, however, that only propagating modes with $k_\perp \le k$ contribute
to the imaginary part of the Greensfunctions,
one can obtain an estimate of the effect in the 
short-wavelength or ray-optics 
limit $(d\gg\lambda)$.

\begin{figure}[htb] 
\begin{center}
    \epsfig{file=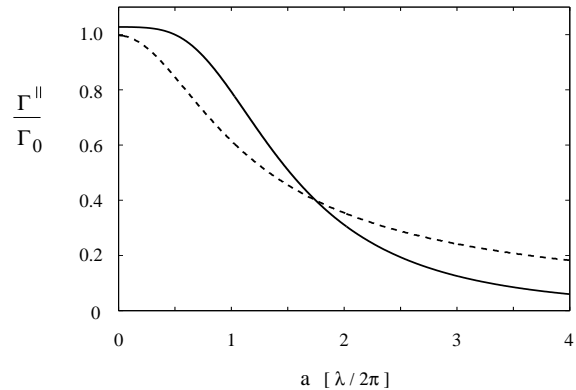,width=7.5cm}
    \caption{$\Gamma^\parallel/\Gamma_{0}$
as function of transversal radius $a$ of the LHM with 
thickness $d=3\frac{\lambda}{2\pi}$. The mirror 
itself was assumed to be infinitely extended (solid) or of the same 
dimension as the LHM (dashed).}
    \label{transversemirror} 
  \end{center} 
\end{figure}

For the system with the mirror this means, that for the integrand 
over $k_\perp$ 
in the definition of the 
Greensfunction (\ref{Green00}) one should use the expression for 
$\Gb^{00}_\text{LHM}$ only for values
\be
k_\perp \le k \frac{\frac{a}{d}}{\sqrt{1 +\left(\frac{a}{d}\right)^2}}.
\ee
This corresponds to angles $\alpha$ of the
propagating modes less than
$\sin\alpha =\dfrac{a}{\sqrt{a^2+d^2}}={k_\perp}/{k}$ 
(fig. \ref{Pendry3D}).
For greater angles the result depends strongly on whether or not the
mirror has also a finite transversal extension. 
When the mirror has the same transversal radius $a$ one has to use
 $\Gb_\text{vac}$ (dashed line fig. \ref{transversemirror}),
otherwise the expression for $\Gb^{00}_\text{LHM}$ needs to be taken, 
but $n=-1$ being substituted by $n=1$,
(solid line fig. \ref{transversemirror}).
%
%
%
%
%
%

\begin{figure}[htb] 
  \begin{center} 
    \epsfig{file=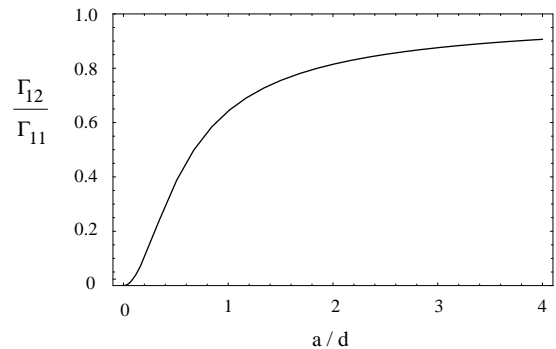,width=7.5 cm}
    \caption{$\Gamma_{12}/\Gamma_{11}$
as function of the normalized transversal radius $a/d$.}
    \label{transverse} 
  \end{center} 
\end{figure}

For the system with the lens an estimate of the effect of a finite 
transverse radius is given here only for a symmetric setup, i.e. the 
distance of both atoms to the surface of the lens being $d/2$.
Under this assumption the sought result can be obtained easier
than for the mirror case, since the atoms are
macroscopically separated. In this case the 
Greensfunction $\Gb_\text{vac}(\rb_1,\rb_2)$ 
is essentially zero and the integration over $k_\perp$ can
effectively be limited to values
\be
k_\perp \le k \frac{\frac{a}{d}}{\sqrt{\frac{1}{4} +\left(\frac{a}{d}\right)^2}}
\ee
with the integrand being the usual expression 
for $\Gb^{20}(\rb_1,\rb_2)$. As can be seen from
figure \ref{transverse}, even for a moderate ratio $a/d$ 
a close to 100\% sub/superradiance 
can be obtained.
%
%
%
%
%
%


\subsection{Dispersion effects}


The spontaneous emission is given by the imaginary part of the retarded
Greenstensor at the transition frequency (\ref{Gamma}). Therefore the 
predictions of sections \ref{pmirror} and \ref{plens}
hold as long as the frequency range with $n=-1$ is large compared
to the natural linewidth $\Gamma$.

\begin{figure}[htb] 
  \begin{center} 
    \includegraphics[width=7.5 cm]{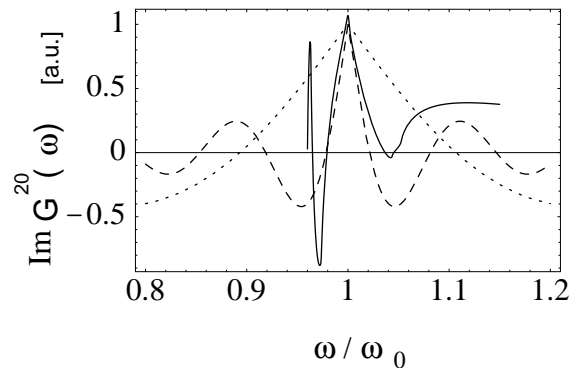}
    \caption{Im$[\Gb^{20}(\omega)]$ 
following from eq.\eqref{Green-disp}
for lossless LHM with $n=-1 +\alpha(\omega-\omega_0)$ for 
$\alpha=45/\omega_0$ for
$dk_0=1$ (dashed), $0.2$ (dotted).
Also shown is the numerically calculated spectrum for a specific 
causal model for $n(\omega)$ with resonances of
$\varepsilon(\omega)$ and $\mu(\omega)$ below $\omega_0$. 
$n(\omega)$ was chosen such that 
 Re$[n(\omega_0)]=-1$ and $\alpha=45/\omega_0$.
The central structure is well represented by the
linear-dispersion approximation \eqref{Green-disp}.
Furthermore a narrowing of the spectral width with increasing thickness
is apparent.}
    \label{fig4} 
  \end{center} 
\end{figure}

The previous discussion suggests, that if the LHM used for the lens in sect.
\ref{plens}
has arbitrarily small losses in the frequency range of interest 
and also has a sufficient large transversal
extension,
sub- and superradiance is possible for two atoms at arbitrary distance.
For causality reasons this is of course not possible. The solution of this
seeming contradiction lies in the necessary dispersion of a 
left-handed material as discussed in sect.II.
The positivity of the electromagnetic energy in a lossless LHM requires that
$\frac{{\rm d}}{{\rm d}\omega} \Bigl(\omega {\rm Re}
\left[\epsilon(\omega)\right]\Bigr) \ge 0$, and 
$\frac{{\rm d}}{{\rm d}\omega} 
\Bigl(\omega {\rm Re}\left[\mu(\omega)\right]\Bigr)\ge 0$,
which implies for $n(\omega_0)=-1:$
\begin{equation}
\frac{{\rm d}}{{\rm d}\omega}n(\omega_0)\ge \frac{1}{\omega_0}.
\end{equation}
As a consequence of the dispersion of the refractive index, 
the frequency window $\Delta\omega$ 
over which $\Gb^{20}(\omega)\approx \Gb^{00}(\omega)$ 
narrows with increasing thickness of the lens. When $\Delta\omega$
becomes comparable to the natural linewidth  of the atomic 
transitions $\Gamma_{11}$, the Markov approximation implicitly used 
for the derivation of eq.\eqref{Liouville} is no
longer valid. To give an estimate when this happens, 
we note from eqs.\eqref{Green20}-\eqref{RS} 
that for $d\gg\lambda$ the term in $\Gb^{20}$
 that is most sensitive to 
dispersion is the exponential factor 
${\rm e}^{i\Kb\cdot(\rb-\rb^\prime)}{\rm e}^{i(k_{1z}-k_z)d}$. 
Taking into account a linear dispersion of $n(\omega)$ 
in this exponential factor, 
according to $n=-1 +\alpha (\omega-\omega_0)$, with a real value of $\alpha$, 
while keeping  the resonance values for $T^{\rm  TE}, T^{\rm TM}$
and  $R^{\rm TE}, R^{\rm TM}$, one finds 
for the Greens-tensor
\be
{\rm Im}[\Gb^{20}(\omega)]=\frac{k}{8\pi}
{\rm Re}\left[\int_0^1 \!\!{\rm d}\xi\, (1+\xi^2) {\rm e}^{i 
\frac{d k_0}{\xi}\alpha (\omega-\omega_0)}\right]\mathbf{\hat 1}.
\label{Green-disp}
\ee
As can be seen from Fig. \ref{fig4} the spectral width $\Delta\omega$ of 
the Greensfunction is in this approximation
of order 
\begin{equation}
\Delta\omega \approx (k_0 d \alpha)^{-1}.
\end{equation}
Since as mentioned above for a lossless LHM $\alpha \ge 1/\omega_0$, one 
arrives at 
\begin{equation}
\Delta\omega\le {c}/{d}.
\end{equation}
This leads to an upper bound
for the distance of the atoms. The requirement $\Delta\omega\gg \Gamma_{11}$
leads to 
\begin{equation}
d \ll \frac{c}{\Gamma_{11}}.
\end{equation}
This condition can easily be understood. It states that the distance
between the two atoms must be small enough such that the travel time
of a photon from one atom to the other is small compared to the 
free-space radiative lifetime. 
%
%
%
%
%
%


\section{Summary}


In the present paper we have studied the interaction of an isolated atom
or a pair of atoms with the quantized electromagnetic field in the
presence of media with negative index of refraction. An expression for the
rate of spontaneous emission of an atom embedded in a LHM was derived
(see also \cite{Dung-PRA-2003}) 
which is a generalization of the Glauber-Lewenstein
result \cite{Glauber} to magneto-dielectric media. We have shown that the
negative optical path length occurring in left-handed materials can be used
to induce strong QED effects over large distances, 
which in vacuum occur only on sub-wavelength length scales. 
Considering an isolated atom
in front of a perfect mirror with a layer of LHM of thickness $d$ we found
an interesting modification of the Purcell effect. Spontaneous emission
was found to be completely suppressed for an atom placed at distance 2$d$
from the mirror in vacuum. It was shown furthermore that two atoms in the
focal points of a Veselago-Pendry lens, consisting of a parallel slab of ideal
LHM, display perfect sub- and superradiance. A principle limitation of 
the involved length scales is given only by the intrinsic dispersion of left
handed materials which prevents the strong radiative coupling over distances 
larger than the propagation distance of light corresponding to the 
free-space radiative decay time. We anticipate that the unusual property of 
LHM to lead to negative optical path length will have a number of interesting 
applications an example being zero-optical length resonators. On the other
hand much of the present discussion is still only of academic interest
since until now no low-loss negative index materials are known for 
the interesting case of optical frequencies.


\def\etal{\textit{et al.}}


\section*{Information about corresponding author}


\

\

M. Fleischhauer 

\

Fachbereich Physik

Technische Universit\"at Kaiserslautern

D-67663 Kaiserslautern

\

phone: ++ 49 631 205 3206

\

fax: ++ 49 631 205 3907

\

email: mfleisch@physik.uni-kl.de
 

\newpage

\end{document}